\documentclass[
aps,%
12pt,%
final,%
notitlepage,%
oneside,%
onecolumn,%
nobibnotes,%
nofootinbib,%
superscriptaddress,%
centertags]%
{revtex4}
\usepackage{pstricks}
\usepackage{fancybox}
\usepackage{xspace}

\begin{document}
\selectlanguage{english}

\title{Proton-nucleus collisions at LHC energy in the Monte Carlo model}
\author{\firstname{V.~N.}~\surname{Kovalenko}}
\email{nvkinf@rambler.ru}
\affiliation{Saint Petersubrg State University,
Russia
}

\begin{abstract}
A Monte Carlo model, initially developed for soft pp and AA collisions at high energy, is
applied for proton-lead interaction at the LHC energy. Elementary collisions are implemented at the
partonic level and do not involve the usual Glauber's supposition of independent nucleon-nucleon
collisions.
The average number of participating nucleons and charged multiplicity in p-Pb collisions
were calculated and compared with the predictions of Glauber model and experimental data. It was
demonstrated that taking into account the energy conservation results in the number of participating
nucleons considerably lower than in the Glauber approach.
Different ways of centrality in determination in pA are discussed and the influence of the
methods of centrality definition on mean observables and their variances is studied.
\end{abstract}

\maketitle


\section{Introduction}

After three years of successful operation of the LHC as proton-proton and lead-lead collider
the first pilot run of proton-lead collisions was held on 12-13 September 2012 \cite{Alemany:1496101}, followed by longer one-month
running period in 2013. The statistics, collected
at the first run, was enough to provide physical results \cite{pPbAliceMult,pPbAliceRPPb,pPbAliceLRC}. 
The first measurements, performed at the LHC, are related to the multiplicity
studies and involve also centrality selection \cite{pPbAliceLRC}.
In the present work two aspects, related to these studies, multiplicity
and centrality, are discussed within the Monte-Carlo model \cite{KovalenkoYad,KovalenkoPoS} and further predictions 
are made.

The first result, obtained by the ALICE Collaboration -- mean pseudorapidity density of
charged particles in NSD p-Pb collisions at 5.02 TeV -- found at the level of $16.81\pm0.71$. 
 The normalization of the result
to the number of participants ($N_{\text{part}}$) was performed and the corresponding normalized value
$2.14 \pm 0.17$ was found to be less than the multiplicity in pp collisions at the same energy (which is
$2.6$, see fig. 2 in \cite{pPbAliceMult}).
Note that the number of participant nucleons was not extracted from the data, but calculated using Glauber model \cite{GlauberFull} (similarly to RHIC experiments), while 
alternative experimental approaches was used
by earlier experiments, such as the
measurement of the number of net
baryons \cite{pApaperNA35} or
ZDC and Veto calorimeters \cite{Alt07p064904}.

The use of Glauber model is criticized  \cite{feofivanov,ASerSP,KovalenkoPoS}, as it over-predicts
the multiplicity yields in AA
collisions and the consistency
can be achieved only by parameters fitting, different
at each colliding system and energy.
In several models behind Glauber the decrease of
the number of binary nucleon collisions ($N_{\text{coll}}$) in AA compared to Glauber model was found \cite{feofivanov,KovalenkoPoS}.
Similar effects are predicted also by the Gribov-Glauber approach \cite{Tywoniuk:2006fq}
and models with gluon shadowing \cite{Deng:2010xg}.
Due to the fact that in pA collisions $N_{\text{part}}=N_{\text{coll}}+1$, the value of
$N_{\text{part}}$ could be also affected by this issues,
and we check this in the present model.
\section{Monte Carlo model}
The present model is based on the partonic picture of nucleons interaction.
The initial positions of the nucleons are distributed
according to Woods-Saxon: $\rho(r)=\frac{\rho_0}{1+exp[(r-R)/d]},$
with parameters $R=6.63$ fm, $d=0.545$ fm.
Each nucleon consist of a set of partons (valence quark, diquark and
quark-diquark pairs), distributed in transverse plane with Gauss distribution relative to the center of nucleon with mean-square radius $r_0$.
For each parton the appropriate momentum fraction is assigned, according
to the exclusive distributions \cite{KovalenkoYad}, accounting energy
and angular momentum conservation in the initial state of a nucleon.

Quark-diquark and quark-antiquark pairs form set of dipoles. Interaction
probability amplitude of two dipoles with transverse coordinates
 $( \textbf{r}_1 \textbf{r}_1')$ and $(\textbf{r}_2  \textbf{r}_2')$ 
 is given by \cite{Lonnbland1,Gustafson}:
\begin{equation} \label{withlog} 		\nonumber
	f=\frac{\alpha_S^2}{8}\ln^2 \frac {( \textbf{r}_1 - \textbf{r}_1' )^2 ( \textbf{r}_2 - \textbf{r}_2' )^2 }
									{( \textbf{r}_1 - \textbf{r}_2' )^2 ( \textbf{r}_2 - \textbf{r}_1' )^2 },
\end{equation}
Note, that two dipoles interact more probably,
if the ends are close to each other,
and (others equal) if they are wide.
After taking into account confinement effects, one gets \cite{Lonnbland1,Gustafson}:
		\begin{equation} \label{newformula}
		\nonumber
			f=\frac{\alpha_S^2}{2}\Big[ K_0\left(\frac{| \textbf{r}_1- \textbf{r}_1'|}{r_{\text{max}}}\right) +
			K_0\left(\frac{| \textbf{r}_2- \textbf{r}_2'|}{r_{\text{max}}}\right) 
			- K_0\left(\frac{| \textbf{r}_1- \textbf{r}_2'|}{r_{\text{max}}}\right)
			- K_0\left(\frac{| \textbf{r}_2- \textbf{r}_1'|}{r_{\text{max}}}\right)	\Big]^2,
		\end{equation}		
where $K_0$ is modified Bessel function. $a_S$ here is an effective coupling 
constant, $r_{\text{max}}\simeq0.2-0.3 \text{ fm}$ -- confinement scale, the exact values are turned to describe experimental data.




The charged multiplicity is calculated in the approach of
colour strings, taking into account their finite rapidity
width and interactions due to non-zero transverse radius of string $r_{\text{str}}$
 (string fusion) \cite{braun11,Braun:2003fn}, with introducing a lattice in the transverse plane \cite{diskr1, diskr2}.

Important feature of the present model is that 
every parton can interact with other one only once, forming
a pair of quark-gluon strings, hence, producing particles, contrary
to Glauber supposition of constant nucleon cross section. A nucleon is participating in the collision
if at least one of it's partons collides with other from the
proton.

Parameters of the model are constrained from the pp data on
the total inelastic cross section and charged multiplicity in wide energy range
(from ISR to LHC) \cite{Khachatryan:2010us,Antchev:2013iaa}.
Additional requirement was the consistent description of the multiplicity
in minimum bias p-Pb and $\text{Pb-Pb}$ collisions at the LHC energy \cite{pPbAliceMult,Aamodt:2010cz}. The remaining freedom in the parameters
selection is used as the systematic uncertainty of the results.

Most of the results were obtained with the following parameters:
\begin{equation}		\nonumber
  r_0=0.6\text{ fm}, \hspace{0.5cm}  \alpha_{S}=0.4, \hspace{0.5cm} r_{\text{max}}/r_0=0.5, \hspace{0.5cm} r_{\text{str}}=0.2\text{ fm}, \hspace{0.5cm}\mu_0=1.152,
\end{equation}
where $\mu_0$ is mean charged multiplicity from one single string per rapidity unit.


\section{Results}
\subsection{$N_{\text{part}}$ and multiplicity}
\begin{figure}[h!]
\begin{center}
\includegraphics[width=0.45\linewidth]{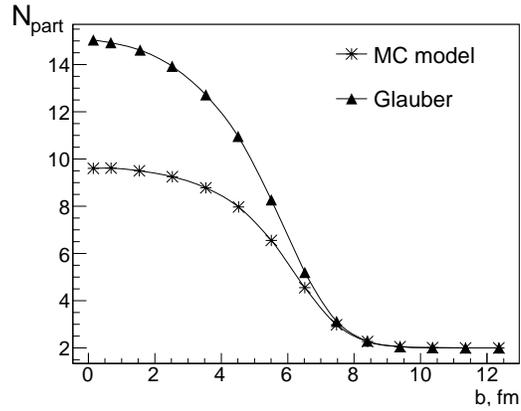}
\caption{Mean number of participant nucleons as a function of impact parameter. Results of the present MC model and Glauber model for p-Pb collisions at $\sqrt{s}=5.02$~TeV.}
\label{fig_1}
\end{center}
\end{figure}

\begin{figure}[h!]
\begin{center}
\includegraphics[width=0.45\linewidth]{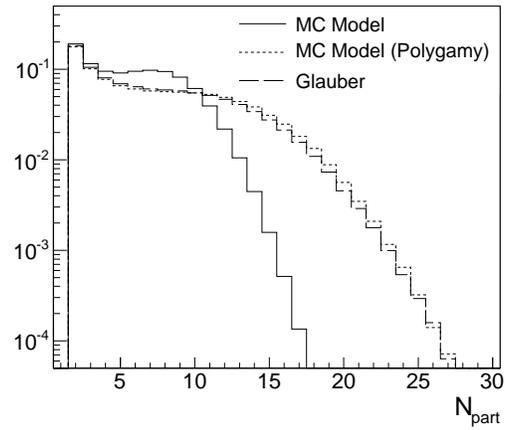}
\caption{Distribution of $N_{\text{part}}$ in p-Pb collisions
at $\sqrt{s}=5.02\text{ TeV}$, calculated in the default, ``polygamous'' MC model and Glauber model.}
\label{fig_2}
\end{center}
\end{figure}

In the framework of the present model we performed the calculations
for p-Pb collisions at $\sqrt{s}=5.02$ TeV.
The mean number of the participant nucleons
in p-Pb collisions is shown at Figure \ref{fig_1} and
compared to the calculations in Glauber model.
The distribution of $N_{\text{part}}$ is shown at Figure \ref{fig_2}.
$N_{\text{part}}$ in the present model is found to be considerable less than in Glauber case,
forming a plateau at central collisions.

In order to clarify the reason of this difference, we implemented
so-called ``polygamous'' version of the model. In this
artificial variant
we allowed the partons to interact several times, forming
the strings and produce particles. Note, that in this case
the same energy, belonging to a parton, can go to the particle 
production several times, breaking the energy conservation.

The number of participants in ``ploygamy'' model (Fig. \ref{fig_2}) is found very close to the Glauber results.
This demonstrates that the accounting of the energy conservation is the reason of decrease of $N_{\text{part}}$ in our non-Glauber approach. 

Note, that similar decrease of $N_{\text{part}}$ in pA compared to Glauber
is found also in several other models, that are aimed to
describe consistently pp, pA and AA collisions, such as
Modified Glauber \cite{feofivanov,ASerSP},
where the energy conservation
is implemented in effective way, and AMPT \cite{pAtheorPaper}, which include gluon shadowing and
collective effects.
Also such signatures were found in the experimental studies \cite{pApaperNA35}, 
when the approach, based on the number of net baryons, was used
for the determination of $N_{\text{part}}$: some discrepancy 
between geometrical models and experimental data was observed only for non-symmetrical colliding systems, although the difference was accounted
as systematic error in the data.

Taking all this into account one may suggest to 
pay more attention, or even reconsider the use of Glauber normalization of the multiplicity yields in experimental studies, at least for
non-symmetrical colliding systems.

\begin{figure}[h!]
\begin{center}
\includegraphics[width=0.45\linewidth]{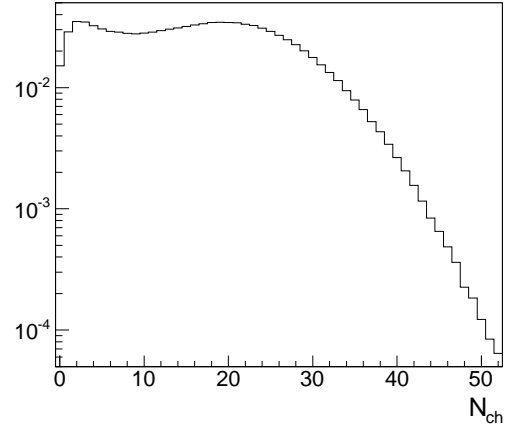}
\caption{Charged multiplicity distribution in the rapidity window |$\eta$|<0.5 for p-Pb collisions at 5.02~TeV, calculated in the MC model.}
\label{fig_3}
\end{center}
\end{figure}

At Figure \ref{fig_3} the prediction of
 multiplicity distribution, calculated
in the present model, is shown. The mean value -- 16.5 -- is consistent
with experiment \cite{pPbAliceMult}. The predicted non-monotonic
shape would be interesting to compare with future measurements.

\subsection{Centrality classes}

We performed also the calculation of centrality dependence
of the multiplicity in p-Pb collisions. Centrality is always
determined as a fraction of some variable among the whole distribution.
We used four centrality estimators: impact parameter, number of participants,
multiplicity signal in off-central rapidity region (``vzero'') and multiplicity in 
$|\eta|<0.5$ itself.
The ``vzero'' signal is 
emulated as a sum of charged multiplicities in rapidity windows: 
(3.0;~5.0)+(-3.6;~-1.6), which is approximately the coverage of the ALICE
detector V0 \cite{06p1295}. We selected five classes of 20\% centrality width.

\begin{figure}[h!]
\begin{center}
\includegraphics[width=0.45\linewidth]{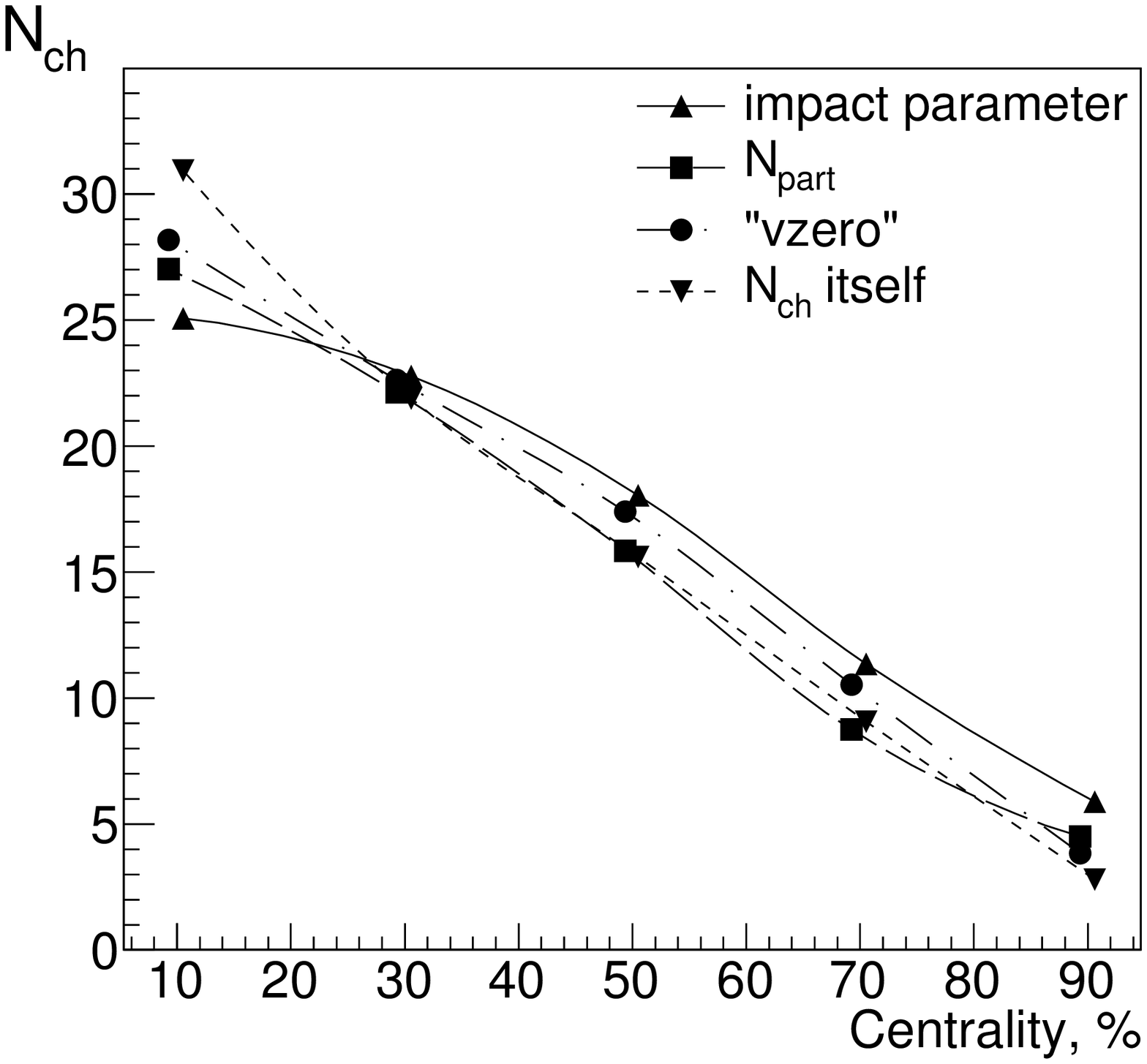}
\includegraphics[width=0.45\linewidth]{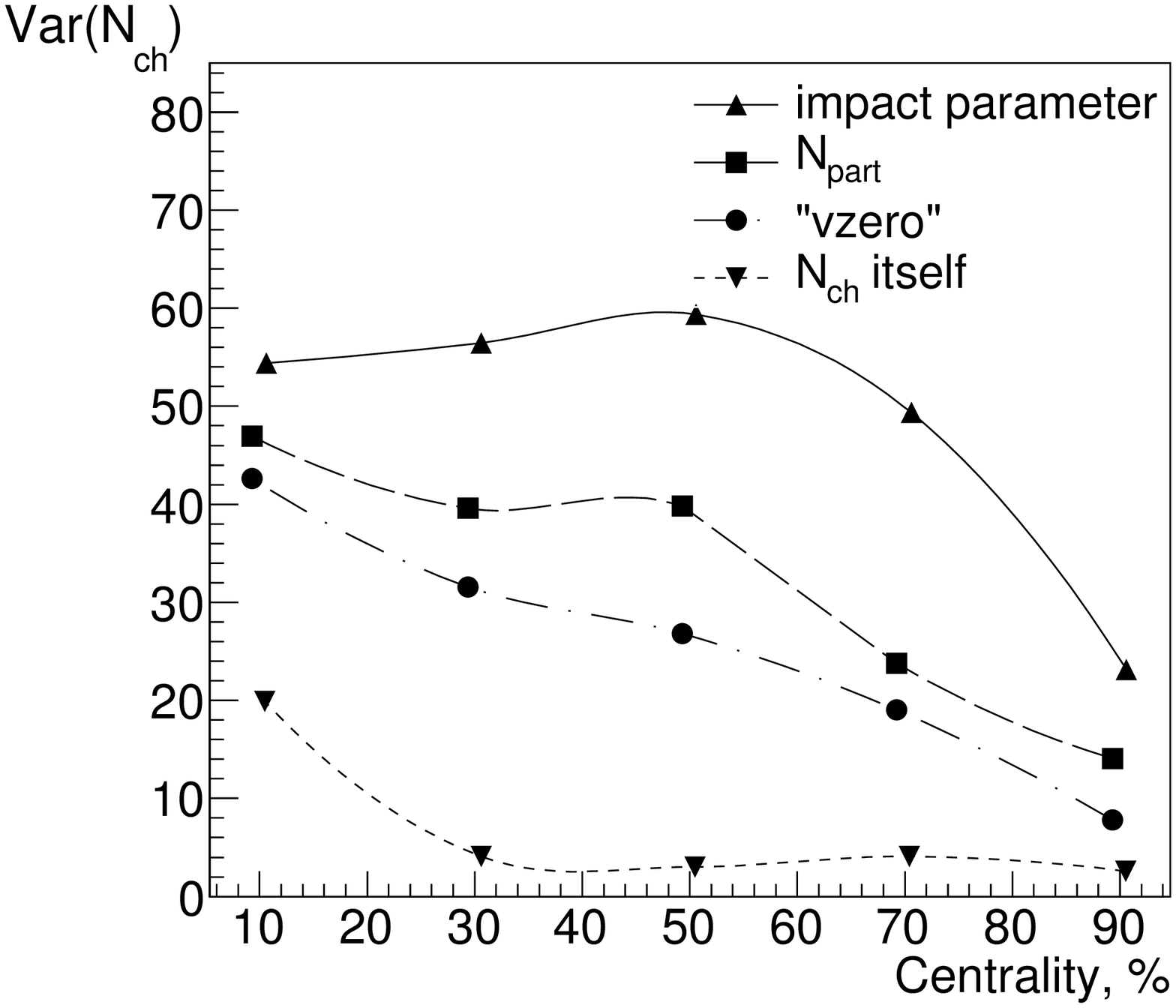}
\caption{Multiplicity in $|\eta|<0.5$ (left) and its variance (right) in p-Pb collisions at 5.02~TeV as function of centrality, obtained in MC model using several centrality estimators: impact parameter, number of participants, multiplicity detector (``vzero'') and multiplicity itself in this window.}
\label{fig_4}
\end{center}
\end{figure}

The results on the multiplicity mean values (Figure \ref{fig_4}) show
noticeable discrepancy between several methods. One concludes that
the difference between central and peripheral collisions
depends on how much the estimator is correlated to the observable.
That leads that
the relation of centrality in pA collisions
to ``geometrical'' properties
is not straightforward, and proper accounting
of the method of the centrality determination
should be performed in order to compare model predictions with experiment
and even between several experiments.

At Figure \ref{fig_4} (right) the variance of the number of charged
particles in several centrality classes is shown. The values and the behavior
is quite different for several centrality estimators.
Assuming
that the model ``vzero'' method would approximate the real ALICE
 experimental centrality
determination, this result can be compared with the further measurements.



\section{Conclusions}

It was demonstrated, that taking into account
the energy conservation in the model
leads to considerable decrease of the mean number of the 
participants  compared to the Glauber model.


The study of several methods of the centrality selection
showed, that there is valuable difference between the
mean multiplicity between methods. Clearly, the way,
how the centrality classes are determined, should be
always taken into account while comparison the results
between several experiments and between models and experiment.

It seems that the ``geometric'' treatment of the centrality is
less relevant in pA collisions, since the true value
of impact parameter is never known in
the real experiment, but relation between it and observables is
not straightforward.


The author would like to thank V.~V.~Vechernin,  G.~A.~Feofilov, T.~A.~Drozhzhova and  A.~Yu.~Seryakov for numerous useful discussions and for motivating the present work.

 \bibliographystyle{maik}
 \bibliography{papers}


%
\newpage

\textbf{Tables}
\begin{table}[h]
\begin{center}
\caption{Summary of the results for NSD p-Pb collisions at 5.02 TeV}
\label{tab:fate}
\begin{tabular}{|l|l|}
\hline
 {MC Model}  & {Glauber and experiment \cite{pPbAliceMult}}\\
\hline
{$\langle N_{\text{part}} \rangle=6.2 \pm 0.6$} & {$\langle N_{\text{part}} \rangle=7.9\pm0.6$} \\
\hline
{$\langle dN/d\eta \rangle=16.5 \pm 0.5$} & {$\langle N_{\text{part}} \rangle=16.81\pm0.71$} \\
\hline
{$ \langle dN/d\eta \rangle / \langle N_{\text{part}}\rangle |_{pA} =2.66\pm0.04$} &
 {$ \langle dN/d\eta \rangle / \langle N_{\text{part}}\rangle |_{pA} =2.14\pm0.17$} \\
\hline
{$ \langle dN/d\eta \rangle / \langle N_{\text{part}} \rangle |_{pp} =2.58$} &
 {$ \langle dN/d\eta \rangle / \langle N_{\text{part}} \rangle |_{pp} =2.58$} \\
\hline

\end{tabular}
\end{center}
\vspace{-0.6cm}
\end{table}

\newpage
{\bf Figures}

\clearpage

\textbf{Figure captions}

\ 

\textbf{Figure 1.} {Mean number of participant nucleons as a function of impact parameter. Results of the present MC model and Glauber model for p-Pb collisions at $\sqrt{s}=5.02$~TeV.}

\ 

\textbf{Figure 2.} Distribution of $N_{\text{part}}$ in p-Pb collisions
at $\sqrt{s}=5.02\text{ TeV}$, calculated in the default, ``polygamous'' MC model and Glauber model.

\ 

\textbf{Figure 3.} Charged multiplicity distribution in the rapidity window |$\eta$|<0.5 for p-Pb collisions at 5.02~TeV, calculated in the MC model.

\ 

\textbf{Figure 4.}
Multiplicity in $|\eta|<0.5$ (left) and its variance (right) in p-Pb collisions at 5.02~TeV as function of centrality, obtained in MC model using several centrality estimators: impact parameter, number of participants, multiplicity detector (``vzero'') and multiplicity itself in this window.

\ 

\

\end{document}